\documentclass{elsart}

\usepackage{amssymb}
\usepackage{amsmath}

\usepackage{graphicx}
\usepackage{graphics}
\usepackage{epsfig}

\journal{New Astronomy}	

\newcommand{\APJ}{Astrophys.\ J.\/}
 
\newcommand{\CP}{Contemp.\ Phys.\/}
\newcommand{\MNRAS}{Mon.\ Not.\ R.\ Aston.\ Soc.\/} 
 
\newcommand{\NJP}{New\ J.\ Phys.\/}
\newcommand{\PL}{Phys.\ Lett.\/} 
\newcommand{\PR}{Phys.\ Rev.\/} 
\newcommand{\PRL}{Phys.\ Rev.\ Lett.\/} 

\begin{document}

\begin{frontmatter}

\title{Higher neutrino mass allowed if Cold Dark Matter and Dark Energy are coupled}

\author[infnmib,unipv]{G. La Vacca},
\ead{lavacca@mib.infn.it}
\author[infnmib,unimib]{S. A. Bonometto},
\author[usc]{L. P. L. Colombo}

\address[infnmib]{I.N.F.N., Sezione di Milano--Bicocca,\\ Piazza della
Scienza 3, 20126 Milano, Italy }
\address[unipv]{Nuclear \& Theoretical Physics Department, Pavia
University,\\ Via Ugo Bassi 3, 22347 Pavia, Italy}
\address[unimib]{Physics Department G.~Occhialini, Milano--Bicocca
University,\\ Piazza della Scienza 3, 20126 Milano, Italy }
\address[usc]{Department of Physics \& Astronomy, University of
Southern California,\\ Los Angeles, CA 90089-0484}

\begin{abstract}
Cosmological limits on neutrino masses are softened, by more than a
factor 2, if Cold Dark Matter (CDM) and Dark Energy (DE) are
coupled. In turn, a neutrino mass yielding $\Omega_\nu$ up to
$\sim0.20$ allows coupling levels $\beta \simeq 0.15\, $ or more,
already easing the coincidence problem. The coupling, in fact,
displaces both $P(k)$ and $C_l$ spectra in a fashion opposite to
neutrino mass. Estimates are obtained through a Fisher--matrix
technique.
\end{abstract}


\end{frontmatter}


\section{Introduction}
\label{sec:intro}
There seem to be little doubt left: at least one neutrino mass
eigenstate or, possibly, two of them exceed $\simeq 0.055~$eV (direct
or inverse hierarchy). This follows solar \cite{solar} and reactor
\cite{reactor} neutrino experiments, yielding $\Delta m_{1,2}^2 \simeq
8 \times 10^{-5}$eV$^2$ and, namely, atmospheric \cite{atmo} and
accelerator beam \cite{beam} experiments yielding $\Delta m_{2,3}^2
\simeq 3 \times 10^{-3}$eV$^2$.

Cosmology is also sensitive to neutrino mass. Since 1984, Valdarnini
\& Bonometto \cite{bono1} made a detailed analysis of transfer
functions in cosmologies where a part of Dark Matter (DM) is due to
massive neutrinos, so proposing mixed DM models, where neutrinos play
an essential role in adjusting CMB (Cosmic Microwave Background)
anisotropies and matter fluctuation spectra to fit observations. A
large deal of work on this subject took place in the Nineties; mixed
models were widely tested, using both the linear and the non--linear
theory.

Hubble diagram of SNIa \cite{bib1} showed then an accelerated cosmic
expansion, while advanced data on CMB \cite{bib2} and large scale
structure \cite{bib3} required a {\it spatially flat} cosmology~with a
matter density parameter $\Omega_{o,m} \simeq 0.27$, so that the gap
up to unity was~to~be~filled by a smooth non--particle component
dubbed Dark Energy (DE).

All that relegated neutrinos to a secondary role in shaping cosmic
data while, by using such advanced astrophysical data, increasingly
stringent limits on neutrino masses could be computed (see {\it e.g.}
\cite{nu1}), also combining cosmological and laboratory data
\cite{fogli}. Moreover, data coming from future weak lensing surveys
seems to be powerful probes of neutrino masses \cite{nulens}.

Standard limits on neutrino masses were recently summarized by Komatsu
et al (2008) \cite{bib30}, within the WMAP5 release, and are quoted in
Table~\ref{tab:nulim}. More stringent but more speculative limits are
suggested in \cite{others1}, who make a more extensive use of 2dF
\cite{bib33} or SDSS \cite{bib34} data, and in \cite{others2}, by
using Ly$\alpha$ forest data.
\begin{table}[b]
\caption{Summary of the 2--$\sigma$ (95\% C.L.) constraints on the sum
of $\nu$ masses, from WMAP 5-year and other cosmological data sets.}
\label{tab:nulim} 
\centering 
\renewcommand{\arraystretch}{0.9}
\begin{tabular}{lcll}
\hline\hline
 && $w = -1$   & $w \neq -1$ \\
\hline
WMAP5        && $< 1.3$~eV  & $< 1.5$~eV   \\
WMAP5+BAO+SN && $< 0.67$~eV & $< 0.80$~eV  \\
\hline
\end{tabular}
\end{table}

These limits, clearly, rely on implicit assumptions concerning the
dark cosmic sector, whose knowledge still fully relies on
astrophysical data, requiring two components characterized by state
parameters $w \simeq 0$ and $\simeq -1$. But the assumption that no
energy exchange between them occurs, tested vs.~data, leads just
to coupling limits.

In this paper we show that spectral distortions due to CDM--DE
coupling and to neutrino mass tend to compensate. We tentatively
estimate how far we can go, simultaneously increasing coupling and
mass, by using a Fisher Matrix (FM) technique. On that basis we
perform a preliminary exploration of the parameter space.
 
A large deal of work dealt with the coupling option (see, {\it e.g.}
\cite{coupling,claudia,maccio,interaction}).  One of its motivations
is the attempt to overcome the {\it coincidence} paradox (see
\cite{chamaleon}), {\it i.e.} the fact that DE becomes relevant just
at the eve of structure formation.  All that makes our epoch unique
and, unless one indulges to {\it anthropic} views, apparently requires
an explanation. However, also independently from this conceptual
issue, our very ignorance of the physics of the dark sector requires
that all reasonable options consistent with basic physics and data are
explored.

It is also important to outline that neutrino mass limits can be
softened if DE with a state parameter $w < -1$ is considered
\cite{Hannestad:2005gj,DeLaMacorra:2006tu}. Unfortunately, this kind
of state equations, yielding the so--called {\it phantom--DE}, can be
justified only making recourse to unconventional physics.

Here, starting from dynamical DE, we shall preliminarily discuss how a
CDM--DE coupling can lead to a context similar to phantom DE. In this
framework, however, no unconventional physics is involved; on the
contrary, thanks to coupling, coincidence could be eased.

For the sake of definiteness, in this paper we use the potential
\begin{equation}
V(\phi) = (\Lambda^{\alpha+4}/\phi^\alpha) \exp(4\pi\, \phi^2/m_p^2)
\label{sugra}
\end{equation}
admitting tracker solutions.  This potential has been shown to fit WMAP
data at least as well as $\Lambda$CDM \cite{bib6} and takes origin
within the context of SUGRA theories \cite{bib5}.

The plan of the paper is as follows. In Section \ref{sec:couplde} we
shall review coupled DE (cDE) models, comparing some aspects of its
physics to {\it phantom }--DE. In Section \ref{sec:spectra} we show
some spectra of a number of cosmological models, showing how CDM--DE
coupling and non--vanishing neutrino mass can be selected so to
(approximately) compensate their effects. In Section \ref{sec:fisher}
we debate technical aspects of a FM approach.  In Section
\ref{sec:fishresult} we give the result of such approach. Section
\ref{sec:explore} is then devoted to a guided exploration of the
parameter space, while in Section \ref{sec:concl} we present our
conclusions.

\section{Coupled--DE models}
\label{sec:couplde}

The essential feature of the scalar field $\phi$, in order that it
yields DE, is its self--interaction through a potential $V(\phi)$. The
simplest form of possible coupling is a linear one. It can be formally
obtained by performing a conformal transformation of Brans--Dicke
theory (see e.g. \cite{brans}), where gravity is modified by adding a
$\phi R$ term ($R$ is the Ricci scalar) to the Lagrangian.

Interactions with baryons are constrained by observational limits on
violations of the equivalence principle (see, {\it e.g.}
\cite{darmour1}) No similar constraints hold for CDM--DE
interactions. In this case, constraints will follow from cosmological
observations.

In the coupled DE (cDE) scenario, as for dynamical DE, a
self--interacting scalar field $\phi$ yields a cosmic component which
does not cluster and has negative pressure. As a matter of fact, its
energy density and pressure read
\begin{equation}
\rho = \rho_k + V(\phi)~,~~~ p = \rho_k - V(\phi)~,
\label{rhop}
\end{equation}
where $V(\phi)$ is the self--interaction potential and
\begin{equation}
\rho_k = \dot \phi^2/2a^2~.
\end{equation}
Here dots indicate differentiation in respect to $\tau$, while the
background metrics reads
\begin{equation}
ds^2 = a^2(\tau) \left[ d\tau^2 - d\lambda^2 \right]
~~~{\rm with}~~~d\lambda^2 = dr^2 + r^2(d\theta^2 + cos^2 \theta
\, d\phi^2)~.
\label{metric}
\end{equation}
If $\rho_k \gg V$, the DE state parameter approaches +1 ({\it stiff
matter}) so that DE energy density rapidly dilutes during expansion
($\rho \propto a^{-6}$). In the opposite case $V \gg \rho_k$, the
state parameter approaches --1 and DE allows the observed cosmic
acceleration. In cDE models, an energy transfer occurs from CDM to
DE, so allowing DE to have a non--negligible density since the
matter--radiation decoupling era.  However, $\rho_k$ is then dominant
and the transferred energy is soon diluted. A so--called $\phi$--matter
dominated period then occurs, when CDM density however declines more
rapidly than $a^{-3}$.  The increase of $\dot \phi$ then brings $\phi$
to approach $m_p$ (the Planck mass) and $V(\phi) $ to exceed
$\rho_k$. DE dilution then stops and DE eventually exceeds CDM density.

Within this picture, CDM and DE stress--energy tensors
($T^{(c,de)}_{\mu\nu}$, let their traces read $T^{(c,de)}$)  no longer
obey separate equations; although still being
\begin{equation}
\label{conti0}
T^{(c)~\mu}_{~~~~\nu;\mu} + T^{(de)~\mu}_{~~~\, ~~\nu;\mu} = 0~,
\end{equation}
it ought then to be
\begin{align}
T^{(de)~\mu}_{~~~\, ~~\nu;\mu} =& +C T^{(c)} \phi_{,\nu}\\
T^{(c)~\mu}_{~~~~\nu;\mu} =&- C T^{(c)} \phi_{,\nu}~.
\end{align}
When the metric is (\ref{metric}), these equations yield
\begin{align}
\ddot \phi + 2 {\dot a \over a} \dot \phi + a^2 V'_\phi =& +C a^2 \rho_c \\
\dot \rho_c + 3 {\dot a \over a} \rho_c =&  -C \rho_c \dot \phi
\end{align}
$\rho_c$ being CDM energy density. General covariance requires $C$
to be a constant or to evolve as a function of $\phi$ itself.
Here, instead of $C$, we shall use the dimensionless parameter
\begin{equation}
\label{bdef}
\beta = (3/16\pi)^{1/2} m_p C~.
\end{equation}
Let us then define the coupling function $f(\phi)$ , through the
relation
\begin{equation}
C(\phi) = {d \log f \over d\phi}
\end{equation}
so that CDM energy density scales according to
\begin{equation}
\label{cevol}
\rho_c(a) = \rho_{o,c} (a_o/a)^3 f(\phi)~.
\end{equation}
Then, if we set $\bar V = V + \rho_c~,$ the $\phi$ eq.~of motion takes
the (standard) form,
\begin{equation}
\ddot \phi + 2 {\dot a \over a} \dot \phi + a^2 \bar V'_\phi = 0~,
\end{equation}
as though CDM and DE were decoupled, once the {\it effective}
potential $\bar V$ is used.

The CDM evolution (\ref{cevol}), implying a density decline faster
than in the absence of coupling, together with Eq.~(\ref{conti0}),
implying that $\rho_c+\rho_{de}$ has the same evolution as in the
absence of coupling, means that $\rho_{de}$ scale dependence is
different from what would follow from the state parameter $w =
p_{de}/\rho_{de}$ deducible from the expressions (\ref{rhop}). The
effective behavior, obtainable by using the potential $\bar V$, mimics
a {\it phantom}--like state equation, yielding a DE density increase
with $a$, as we would find for $w < -1~.$

This makes reasonable to expect that neutrino mass limits can be
relaxed in a cDE context, as they are in the presence of phantom DE.
This option, however, does not lead to requiring unconventional
physics. On the contrary, if we are allowed to consider fairly high
$\beta$ values, the {\it coincidence} problem is also eased.

\begin{figure}[t!]
\centering
\includegraphics[height=12.cm,angle=0]{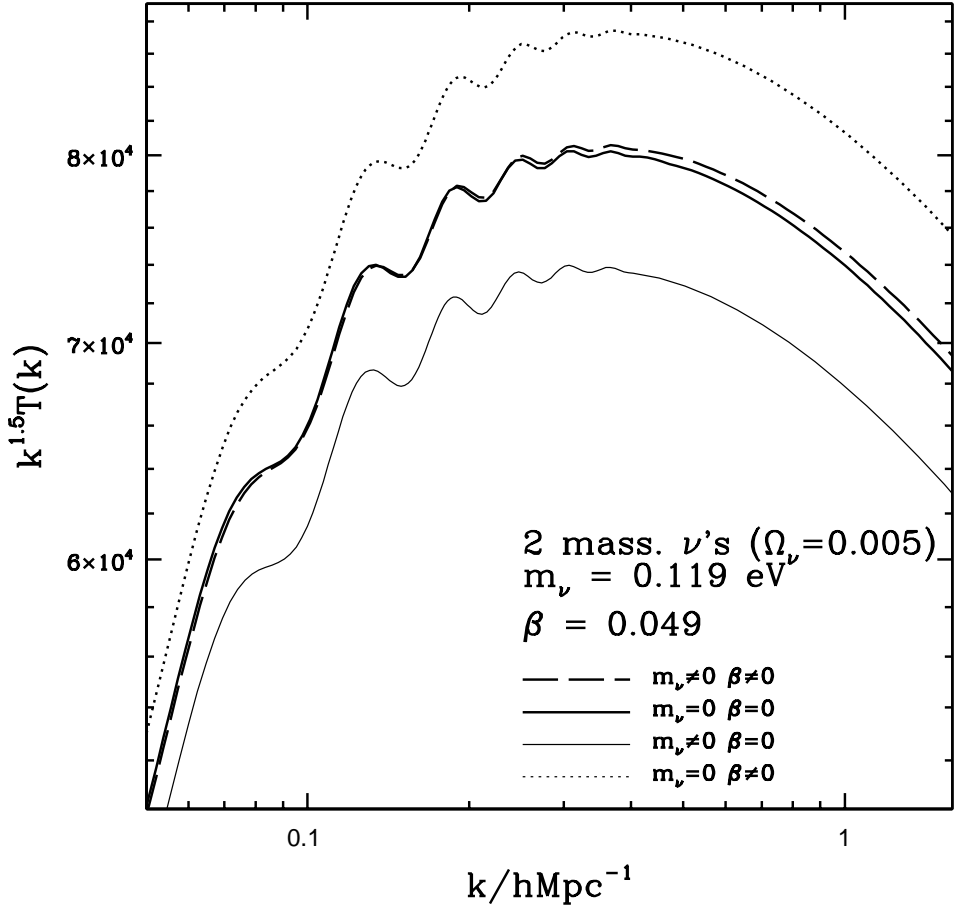}
\caption{\textsl{Transfer functions in cosmologies with/without
coupling and with/without 2 massive neutrinos. Coupling and mass are
selected so to yield an approximate balance. The functions are
multiplied by $k^{1.5}$, to help the reader to distinguish different
cases.}}
\label{t2n}
\vskip +.1truecm
\end{figure}

\section{Some angular and linear spectra }
\label{sec:spectra}
\begin{figure}[h!]
\centering
\vskip-.2truecm
\includegraphics[height=13.cm,angle=0]{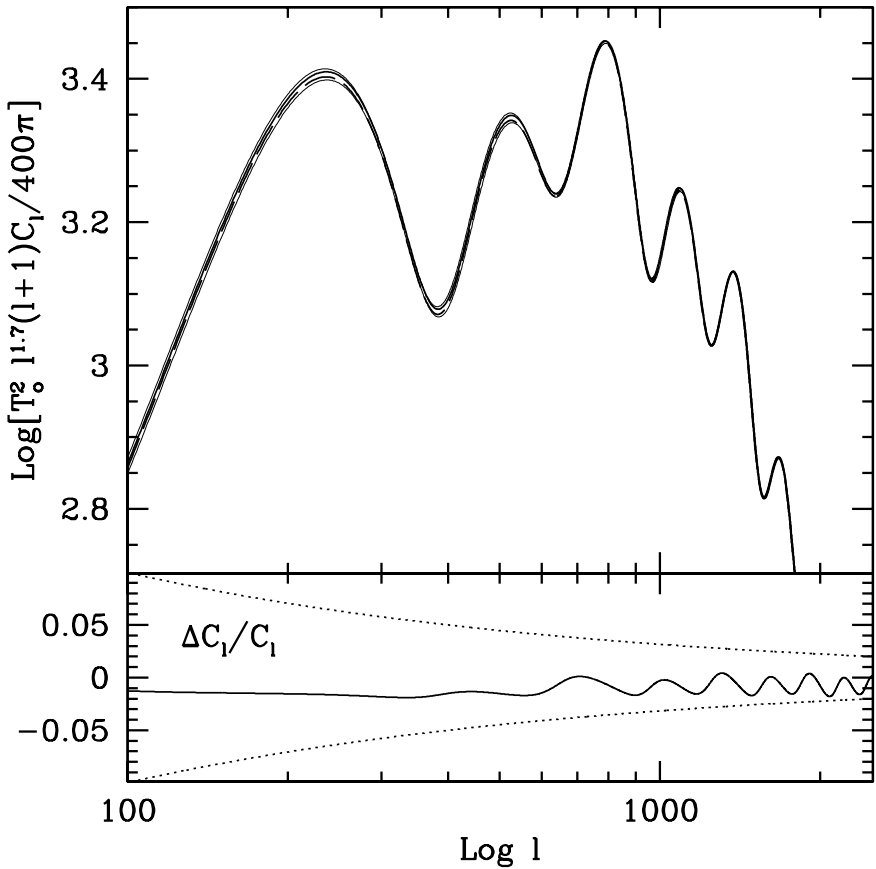}
\caption{\textsl{Angular anisotropy spectra for the same models of the
previous figure. Due to intrinsic $C_l$ oscillations, this Figure is
slightly harder to read. In the lower frame we also give the spectral
differences between 00-- and CM--models. Large $l$ oscillations could
be further damped by a shift by 1 or 2 units along $l$. The dotted
lines represent the cosmic variance interval.  }}
\label{c2n}
\vskip +.1truecm
\end{figure}

The point of this paper can be appreciated through the spectra in
Figures \ref{t2n} and \ref{c2n}. We compare a model with zero coupling
and zero neutrino mass (00--model, hereafter) with: (i) a model with 2
massive neutrinos with mass $m_\nu = 0.119\, $eV, yielding $\Omega_\nu
= 0.005$ (plus 1 massless neutrino); (ii) a model with a CDM--DE
coupling $\beta = 0.049;$ (iii) a model with both neutrino mass and
coupling (CM--model, hereafter).  All models are spatially flat, have
adimensional Hubble parameter $h = 0.71$, density parameters $\Omega_b
= 0.04$, $\Omega_{de} = 0.73$, spectral index $n_s = 0.96$ and a
cosmic opacity $\tau_{opt} = 0.089$. DE is due to a SUGRA potential
with $\Lambda = 1.1\, $GeV, fitting WMAP and other data at least as
well as $\Lambda$CDM. The slope $\alpha$ of the SUGRA potential is
then fixed by requiring that the field density today matches
$\Omega_{de}$.

Angular and spatial spectra are computed with an extension of the
program CAMB \cite{camb}, allowing to treat coupled DE models also in
the presence of massive neutrinos.

DE treatment requires that 4 first order differential equations are
added to the basic budget. In the CAMB implementation used for this
paper we account for the dynamical evolution of DE by using the
variables $\phi(\tau)$ and $\dot \phi(\tau)$ for the background DE
field, as well as the variables $\varphi (\tau,k)$ and $\dot
\varphi(\tau,k)$ for DE fluctuations.

Also CDM dynamics shall be modified when coupling is considered. The
equations, reported in Appendix A, are easily obtainable, {\it e.g.},
from \cite{atv}.

Before running CAMB we need a precursor program, to determine the
value of $\alpha$ consistent with the assigned $\Omega_{de}$ and
$\Lambda$.

Initial conditions for background variables are easily set according
to known tracker solutions. In the presence of coupling, the tracking
regime for density fluctuations is complex and includes different
alternatives. We however found that, if we set $\varphi=0$ and $\dot
\varphi=0$ at an initial time $\tau_{in}$, or we choose their
expressions according to an arbitrary alternative, fluctuations
accommodate rapidly on the right tracking and are completely
independent from the initial choice at a time $\tau(k)$ when the $k$
scale enter the horizon, provided that $\tau_{in} \ll \tau(k)$.

In the linear theory, however, DE fluctuations matter just about the
horizon scale and are rapidly damped afterwards.

Both $l$ and $k$ ranges are selected for being those physically most
significant. At lower $l$'s model discrepancies essentially vanish.
In the $l$ region shown, we have the sequel of maxima and minima due
to primeval compression waves.  The $k$ range covers the scale
explored by deep samples, as 2dF or SDSS, up to $k$ values where
non--linear effects become important.

In the plots, spectra are multiplied by suitable powers of the
abscissa $l$ or $k$, so to reduce the ordinate range. In spite of
that, in the $C_l$ plot different spectra are not easy to
distinguish. We then plot also the ratio $\Delta C_l/C_l$ at constant
$l$; shifts would however appear even smaller if slight shifts along
the $l$ axis (by 1 or 2 units) were performed.

The Figures are principally meant to show that the effects of neutrino
mass and coupling are opposite. The coupling intensity, in fact, is
selected so to (approximately) balance neutrino masses.

We took, however, $\sum m_\nu \ll 0.67$ and $\beta \ll 0.075$ (see
\cite{coupling,claudia,maccio}); each of these values, by itself, is
within current observational limits.  Accordingly, even the difference
between thin and thick solid--line spectra cannot be appreciated
through current data.

In particular, let us outline how the BAO (baryonic acoustic
oscillation) structure is faithfully reproduced when passing from the
00--models (thick solid line) to the CM--models (thick dashed line).

\section{Fisher matrix (data and technique)}
\label{sec:fisher}
We then aim to test how far we can go, simultaneously increasing
$\beta$ and $\Omega_\nu$, without conflicting with data. This can be
estimated by using a FM analysis
\cite{Fisher:1935,Sivia:1996,Tegmark:1996bz}.

This approach allows a rapid, semi--analytic estimate of the
confidence limits for a specific experiment. It assumes a reference
model as the most probable one, {\it i.e.} as the maximum of the
likelihood distribution ${\cal L}({\vec {\rm x}}|\vec\theta)$ of the
data system $\vec {\rm x}$ given the model, described by parameters
$\vec\theta \equiv (\theta_i)$. Exploiting this hypothesis, one can
approximate ${\cal L}$ by a multivariate Gaussian distribution, built
using its second derivatives in respect to the parameters $(\bar
\theta_i)$ at the reference model. Nevertheless, as is known, this
technique is limited by the actual non--Gaussian behavior of data.

FM is nothing but the Hessian of the log-likelihood function:
\begin{equation}
F_{ij} = -\left({\partial \over \partial \theta_i} {\partial \over
\partial \theta_j} \log {\cal L} \right)_{\vec \theta} =
\sum_{\ell\ell'} \frac{\partial {\rm x}_\ell}{\partial \theta_i} {\rm
Cov}^{-1}_{\ell\ell'}(\vec\theta) \frac{\partial {\rm
x}_{\ell'}}{\partial \theta_j} \,.
\label{fisher}
\end{equation}
In the literature, cosmological models are constrained by using a
large number of observables. To our present aims we shall directly
consider the spectrum of matter fluctuations $P(k)$ and the CMB
angular spectra $C_l^{XY}$ ($XY = TT,\, TE,\, EE$). In their recent
analysis, Komatsu et al (2008) made a more restricted use of $P(k)$,
using only BAO's, while they used SNIa Hubble diagrams, so significant
also for being the first signal of DE.

Here we chose observables directly coming from the model, in the
attempt to leave apart observational biases, focusing just on the
level of sensitivity of possible experiments. We consider then two
different experimental contexts. The first one assumes that CMB
spectra are measured at WMAP sensitivity and $P(k)$ is measured with
the sensitivity of the 2dF experiment (case W). The second assumes
PLANCK \cite{bib32} sensitivity for CMB spectra and SDSS sensitivity
for $P(k)$ (case P). The observational features for each mission
considered are listed in Table~\ref{tab:1} for the case of CMB
experiments and the galaxy surveys.
\begin{table}[b]
\caption{CMB (upper table) and galaxy surveys (lower table)
specifications used in the paper. In the lower table, scales and
volumes are in Mpc/h and (Mpc/h)$^3$, respectively.}
\label{tab:1}
\centering
\renewcommand{\arraystretch}{1.1}
\begin{tabular}{lccccc}
\hline 
Mission & $l_{max}$ & $f_{sky}$ & $\theta_{FWHM}$ &
$\sigma_T$ & $\sigma_P$ \\ 
\hline 
WMAP & 1000 & 0.8 & 13' & 260 & 500 \\ 
PLANCK & 2500 & 0.8 & 7.1' & 42 & 80 \\ 
\hline
\end{tabular}
\hskip.1\textwidth
\begin{tabular}{lccc}
\hline 
Mission & $k_{min}$ & $k_{max}$ & Volume \\
\hline 
2dF & 0.02 & 0.1 & $10^8$ \\
SDSS & 0.02 & 0.15 & 0.72$\times10^9$ \\
\hline
\end{tabular}
\end{table}

Let us now consider first the use of CMB data only and let $C_l^{XY}$
be the angular spectra of the input model, to which we must add a
white noise signal, to obtain
\begin{equation}
\bar C_l^{XY} = C_l^{XY} + N_l^{XY}  
~~~~~~
{\rm with}
~~~~~~
 N_l^{XY} = \delta_{XY}\sigma_X^2
 \exp\left[l(l+1)\frac{\theta_{FWHM}^2}{8\ln2}\right]~.
\end{equation}
The expressions of the Fisher matrix $F^{ij}_{C}$ components are then
obtainable according to the relation
\begin{equation}
F^{ij}_{C} = 
\sum_l \sum_{XY,X'Y'} {\partial \over \partial \theta_i} C_l^{XY}
[{\rm Cov}_C^{-1}]^{XYX'Y'}_l
 {\partial \over \partial \theta_j} C_l^{X'Y'}
\label{eq:cmbfisher}
\end{equation}
with
\begin{equation}
[{\rm Cov}_C]_l^{XY,X'Y'} = {1 \over (l+1/2)f_{sky}}
 \left( \begin{array}{ccc}
(\bar C_l^{TT})^2 & (C_l^{TE})^2 & C_l^{TE} \bar C_l^{TT}\\
(C_l^{TE})^2 & (C_l^{EE})^2 & C_l^{TE} \bar C_l^{EE}\\
C_l^{TE} \bar C_l^{TT} & C_l^{TE} \bar C_l^{EE} & \frac{1}2[(C_l^{TE})^2 
+ \bar C_l^{TT} \bar C_l^{EE}]
 \end{array} \right)\,.
\end{equation}
On the contrary, when dealing with matter power spectra, we used the
following definition for the FM \cite{Neyrinck:2006zi} 
\begin{equation}
F^{ij}_P = \sum_{\alpha,\beta} {\partial \over \partial \theta_i}
P(k_\alpha) [{\rm Cov}_P^{-1}]_{\alpha\beta} {\partial \over \partial
\theta_j} P(k_\beta)
\label{eq:mpsfisher}
\end{equation}
with
\begin{equation}
[{\rm Cov}_P]_{\alpha\beta} \simeq
\delta_{\alpha\beta}\frac{V_f}{V_s(k_\alpha)}2P^2(k_\alpha)\,,
\label{eq:mpscov}
\end{equation}
where $V_f=(2\pi)^3/V$ is the volume of the fundamental cell in $k$
space, $V$ is the volume of the survey and $V_s(k_\alpha)=4\pi
k_\alpha^2 \delta k$ is the volume of the shell of width $\delta k$
centered on $k_\alpha$ \cite{Scoccimarro1999,Hamilton2006}. In
Eq.~(\ref{eq:mpscov}) we left aside the contribution of the
trispectrum, because in our analysis we considered only the linear
scales, where the trispectrum is expected to be negligible.

The cosmological model we consider is characterized by 9 parameters:
\begin{eqnarray}
\omega_b = \Omega_{\rm b}h^2 &:\quad& {\rm
Physical\;baryon\;density}\nonumber\\ \omega_c = \Omega_{\rm c}h^2
&:\quad& {\rm Physical\; CDM\; density}\nonumber\\ H_0 &:\quad& {\rm
Hubble\; constant}\nonumber\\ A_s &:\quad& {\rm Scalar\; fluctuation\;
amplitude}\nonumber\\ n_s &:\quad& {\rm Scalar\; spectral\;
index}\nonumber\\ \tau_{\rm opt} &:\quad& {\rm Reionization\;
optical\; depth}\nonumber\\ {\rm Log}(\Lambda/{\rm GeV}) &:\quad& {\rm
Decimal\; logarithm\; of\; the\; energy\; scale\; in\; SUGRA\;
potential}\nonumber\\ \beta &:\quad& {\rm CDM-DE\; coupling
\;strength}\nonumber\\ \sum m_{\nu}/{\rm eV} &:\quad& {\rm sum\; of\;
neutrino\; masses}\nonumber
\end{eqnarray}
We estimate the neutrino mass density parameter, $\Omega_\nu h^2$,
converting it from the total neutrino mass via
\begin{equation}
\Omega_\nu h^2 = \frac{\sum m_\nu}{93.5\;{\rm eV}}\,.
\end{equation}
We compute the CMB anisotropies (temperature and polarisation) power
spectra and the transfer functions, used to calculate linear matter
power spectrum, using a modified version of CAMB. Double sided
numerical derivatives were evaluated considering a $5\%$ stepsize,
except for $\Lambda$, where we adopted a $5\%$ stepsize on $\lambda
\equiv {\rm Log} (\Lambda /{\rm GeV})$.

\section{Fisher matrix (results)}
\label{sec:fishresult}
\begin{figure}[t!]
\centering
\vskip .2truecm
\includegraphics[height=12.cm,angle=0]{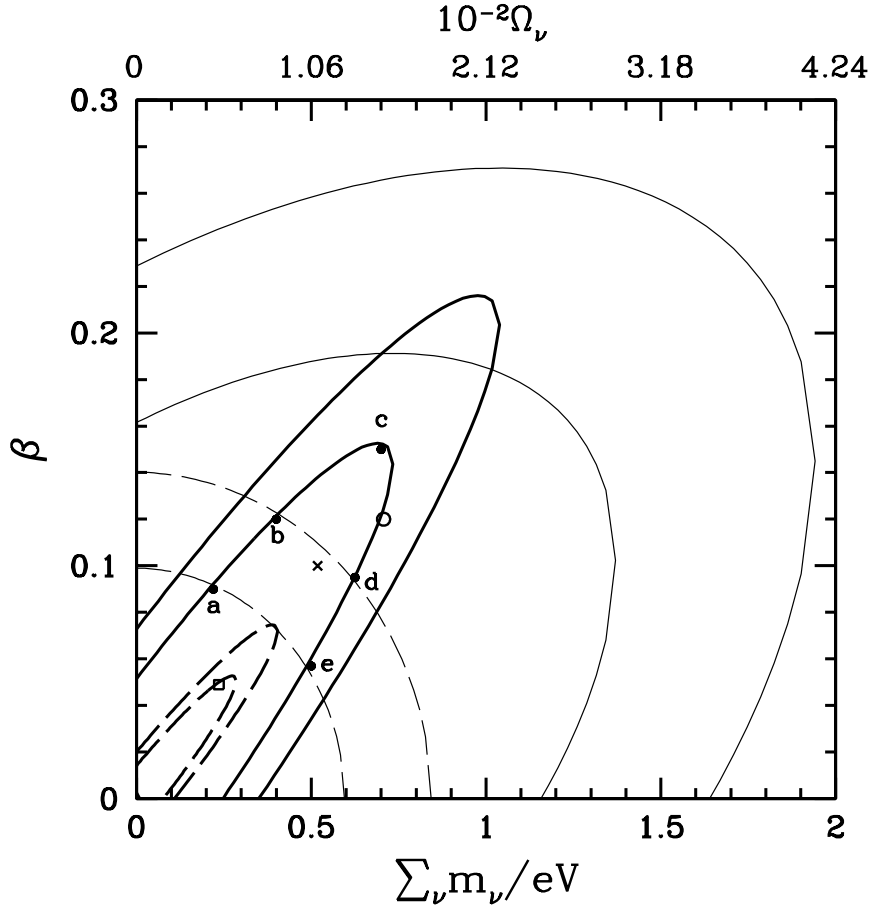}
\caption{\textsl{1-- and 2--$\sigma$ confidence levels for a
2--massive--neutrino model, assuming that the true cosmology is a
SUGRA model with $\log(\Lambda/{\rm GeV})=1.1$, while $\beta = 0$ and
$\Omega_\nu \simeq 0$. Thick (thin) curves show the constraints
deriving from CMB and deep sample data (from CMB data only).  Solid
curves refer to the (i) case (WMAP+2dF). Dashed curves refer to the
(ii) case (PLANCK+SDSS). In the sequel we shall examine in detail
models corresponding to the points labeled a, b, c, d, e and others.
The location of the CM--model of Figs.~\ref{t2n} and \ref{c2n} is
indicated by an open box. The two locations indicated by an open
circle and a cross will also be considered in detail below. This
Figure is somehow analogous to Fig.~2 in Hannestad, 2005. }}
\label{mbmap}
\vskip +.1truecm
\end{figure}

In Figure \ref{mbmap} we then report the expected 1-- and 2--$\sigma$
likelihood curves on the $\sum m_\nu$--$\beta$ plane, for both cases W
and P. In either case we analyse the constraints coming just from CMB
data and those arising from the joint exploitation of CMB and deep
sample data. We performed the analysis either assuming 3 equal mass
neutrinos, or 1 massless and 2 massive neutrinos. The plots shown in
the Figure are obtained for the latter case, but discrepancies are
just a minor effect.
%
\begin{figure}[h!]
\centering
\vskip .2truecm
\includegraphics[height=7.cm,angle=0]{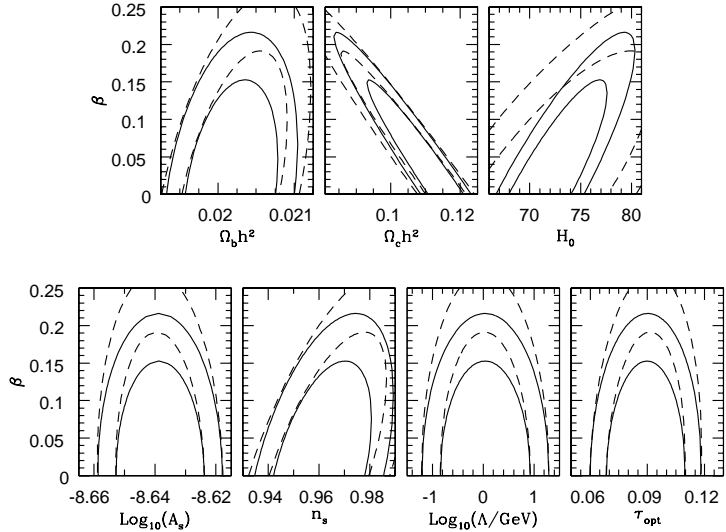}
\caption{\textsl{Correlation between $\beta$ and the other model parameters
for the W case (WMAP+2dF); dashed lines refer to CMB data only.
}}
\label{ballmap}
\vskip +.1truecm
\end{figure}
\begin{figure}[h!]
\centering
\vskip .2truecm
\includegraphics[height=7.cm,angle=0]{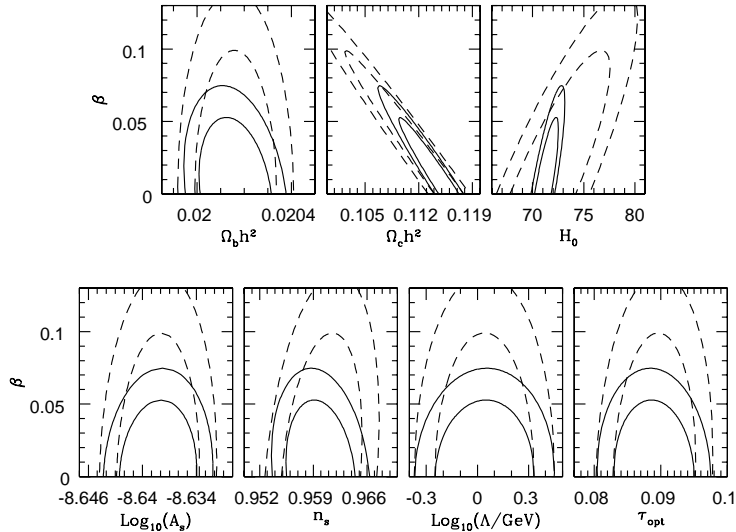}
\caption{\textsl{Correlation between $\beta$ and the other model parameters
for the P case (PLANCK+SDSS); dashed lines refer to CMB data only.
}}
\label{ballpla}
\vskip +.1truecm
\end{figure}
\begin{figure}[h!]
\centering
\vskip .2truecm
\includegraphics[height=12.cm,angle=0]{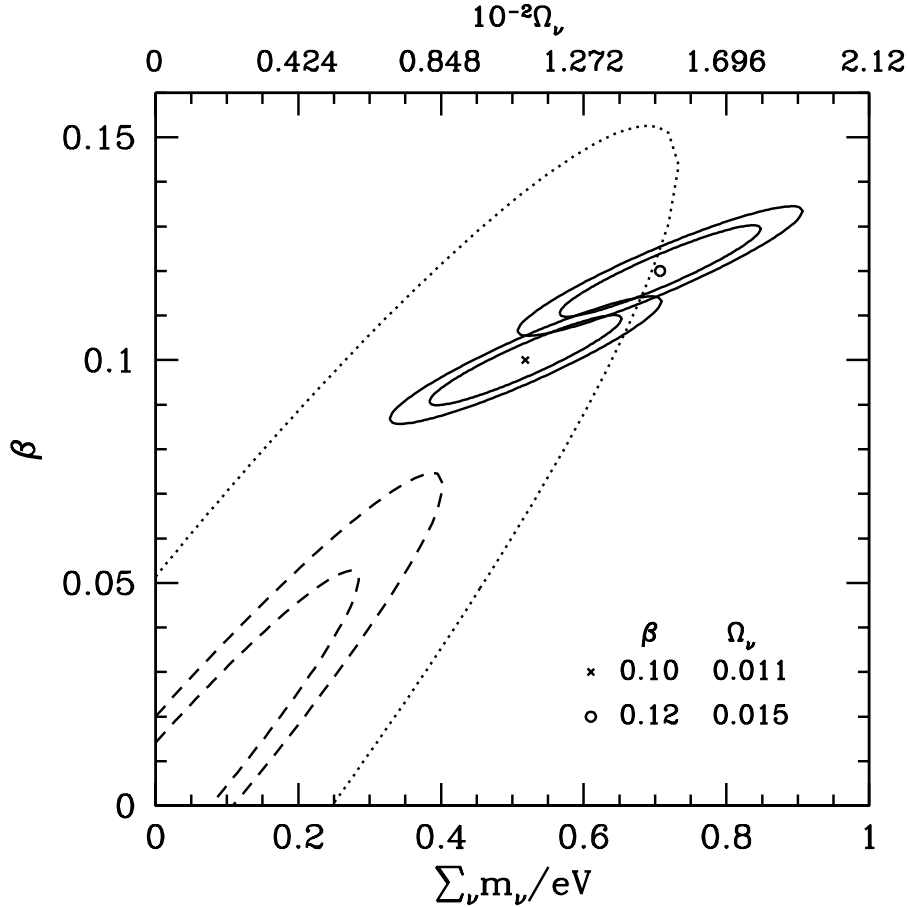}
\caption{\textsl{1-- and 2--$\sigma$ limits, for the P--case, assuming
that the true cosmology corresponds to the points marked by cross or
open circle. The dashed lines report the same limits around the
00--model, as shown in Fig.~\ref{mbmap}; similarly, the dotted line is
the 1--$\sigma$ limit around the 00--model in the W--case, as shown in
Fig.~\ref{mbmap}. This Figure shows that cosmologies, comprising
CDM--DE coupling and neutrino masses, presently compatible with the 00
option, will be easily discriminated, at the P sensitivity level. }}
\label{final}
\vskip +.1truecm
\end{figure}

The Fisher--matrix results, for the W case, substantially confirms
known 1-- and 2--$\sigma$ limits on $\beta$
\cite{claudia,maccio,couplconstr}, yielding $\beta < 0.05$ and $\beta
< 0.075$, respectively, along the $\beta$ axis ({\it i.e.}  with $\sum
m_\nu = 0$).

On the other axis, with $\beta = 0$, $\sum m_\nu$ seems to be more
constrained than what we know from current limits ( $\sum m_\nu <
0.35$~eV vs.~$\sum m_\nu < 0.8$~eV with $w \ne -1$).  These
discrepancies can be read as an indication of the level or reliability
that Fisher--matrix estimate can have. In particular, they may be
partially due to the impact of using the whole $P(k)$ information, as
well as to the fact that the reference cosmology is SUGRA instead of
$\Lambda$CDM. However, the CMB 2--$\sigma$ constraint we find, $\sum
m_\nu < 1.65$~eV, is close to the 95$\, \%$ confidence limit $\sum
m_\nu < 1.5$~eV obtained through a full MonteCarlo analysis of WMAP
data only, with $w \ne -1$.

The likelihood plots have the expected shape. Taken at face value they
yield upper limits $\beta \lesssim 0.22$ and $\sum m_\nu \lesssim
1.05~$eV , in the case W. With the value of $H_o$ used here this would
correspond to $\Omega_\nu \simeq 0.022$, more than 50\% of baryon
density.

On the contrary, in the case P, constraints are more severe, as only
CM--models with $\beta < 0.07$ and $\sum m_\nu < 0.4$~eV appear
consistent with the 00--model, at the 2--$\sigma$ level. These limits
are close to the maximum coupling and neutrino mass separately
admitted in the present observational constraints.

In Figure \ref{ballmap} we also show the correlations between $\beta$
and the whole set of parameters considered, in the W
case. Correlations can be considered negligible for the parameters
$A_s$, $n_s$, $\Lambda$, $\tau_{opt}$ . The correlations with the
parameters $\omega_c$, $\omega_b$, $H_o$, as expected, are stronger.
Figure \ref{ballpla} yields analogous results for the P case.

It may also be useful to consider Figure \ref{final}, showing that
models, including CDM--DE coupling and neutrino masses, compatible with
the 00 option at the W sensitivity level, at the P sensitivity level
will be well discriminated from it and/or also between them.

\section{Exploring the parameter space}
\label{sec:explore}
\begin{figure}[h!]
\centering
\vskip .2truecm
\includegraphics[height=12.cm,angle=0]{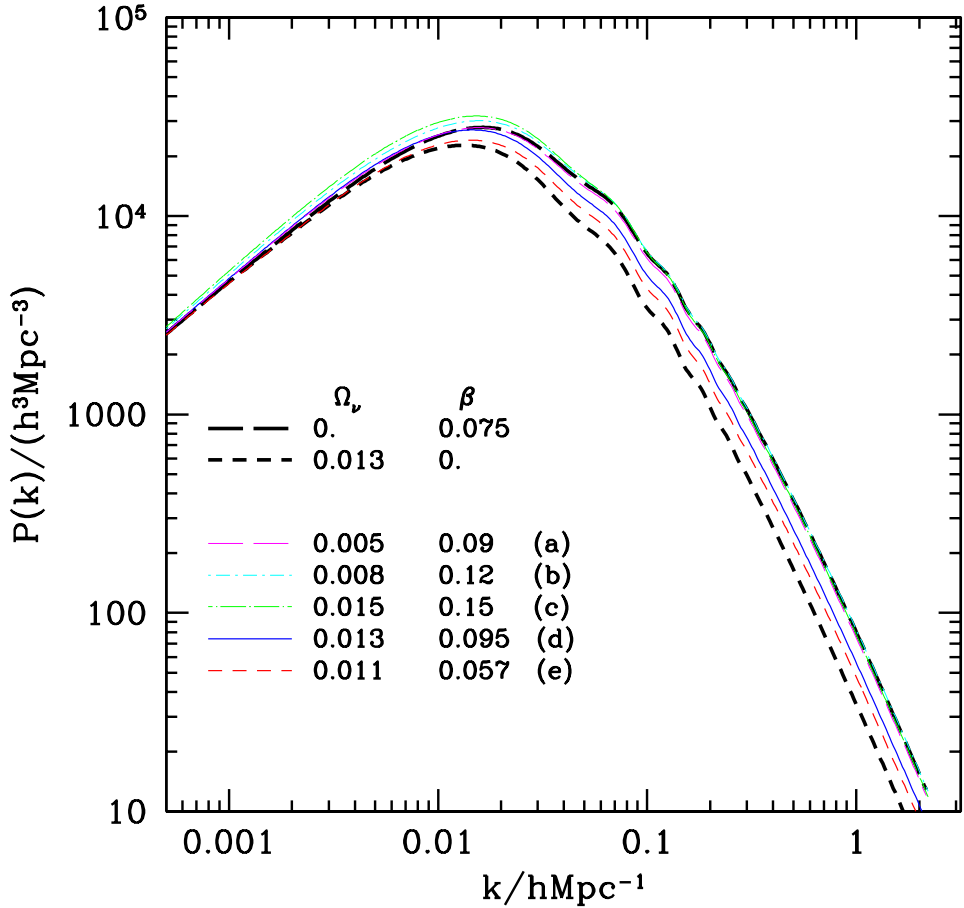}
\caption{\textsl{Spectra for a number of cosmologies. All of them are
obtained setting $n_s = 0.96 $ and $\log A_s = -8.64$, so to enhance
model differences. Thick lines correspond to models presently
considered in agreement with data, and yielding maximum values either
for $\sum m_\nu$ or $\beta$. The other lines yield models
corresponding to the points {\it a, b, c, d, e} in Fig.~\ref{mbmap},
consistent with the 00--model at the 1--$\sigma$ level.}}
\label{spe}
\vskip +.1truecm
\end{figure}

\begin{figure}[h!]
\centering
\vskip .2truecm
\includegraphics[height=13.cm,angle=0]{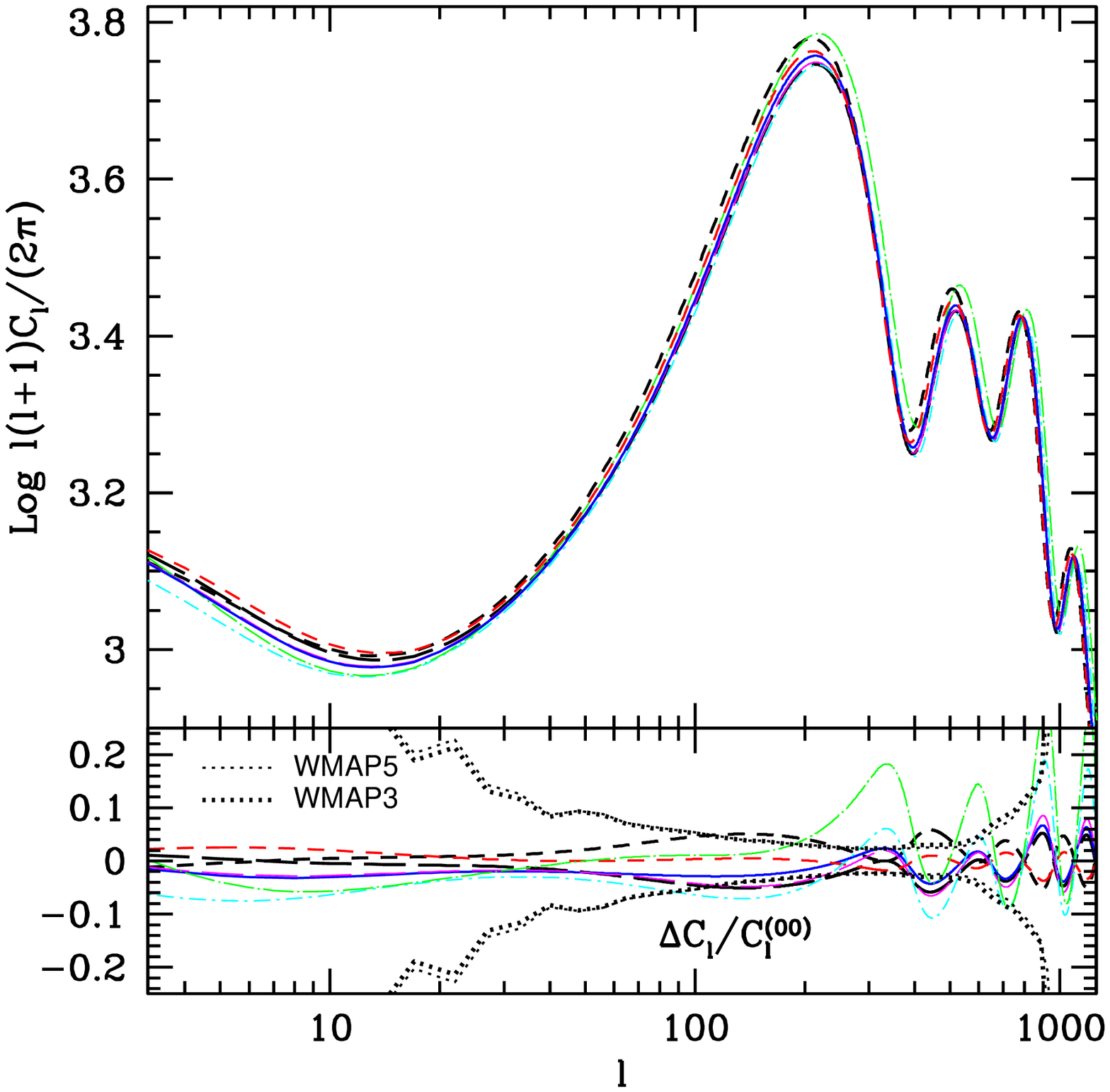}
\caption{\textsl{Spectra of CMB anisotropies for the same cosmologies
of Fig.~\ref{spe}, compared with WMAP error amplitudes. Let us remind
that all of them are obtained keeping the same values $n_s = 0.96 $
and $\log A_s = -8.64$, so to enhance model differences. The relative
difference of the thick line models from the 00--model appears not so
wide as for some of the other models.  Among them, however, the solid
and dashed line models seem to perform quite well. Their performance
can be improved by adjusting the $H_o$ value, slightly modifying
Fisher matrix outputs.}}
\label{clerr}
\vskip +.1truecm
\end{figure}

Of course, a FM analysis gives just a basic idea of the precision with
which current/future data can constrain our parameter set and relax
current bound on $\Omega_\nu$; on the contrary, by no means a FM
analysis can tell us whether data favor $\beta=0$ or $\neq 0:$ here,
when computing the FM from WMAP+2dF--like data, we choose {\it
arbitrarily a fiducial model with $\beta=0$, $\Omega_\nu=0$, assuming
it to have a high likelihood on the $\beta$--$\Omega_\nu$ plane.

While a FM approach allows no likelihood estimate,} in this Section we
wish to provide a few examples, exploring the parameter space that FM
results apparently allow and, in Figures \ref{spe} and \ref{clerr}, we
exhibit the spectra for a set of models. 

As stated in the frame of Fig.~\ref{spe}, the models yielding maximum
CDM--DE coupling (and vanishing $\nu$ mass) or maximum neutrino mass
(and vanishing coupling) have the thick line spectra. Model
discrepancy is enhanced by taking the same amplitude $A_s$, instead of
normalizing them to the same $\sigma_8$.

The setting of models {\it a, b, c, d, e} on the $\sum m_\nu$--$\beta$
plane is indicated in Fig.~\ref{mbmap}.  They are typically within
1--$\sigma$ boundaries. The best performance, perhaps, can be ascribed
to models $d$ and $e.$ Both of them yield a present hot dark matter
density exceeding 1$\, \%$ of the critical density and 5$\, \%$ of the
whole DM.

One of the scopes of this Section was testing that {\it mildly} mixed
DM models, in the presence of a significant CDM--DE coupling, are
reasonably consistent with data.

\section{Conclusions}
\label{sec:concl}
This paper performs a first inspection on the possibility that high
$\nu$ masses and CDM--DE coupling yield compensating distortions of
matter fluctuations and CMB spectra. This compensation is highly
effective for small masses and couplings, as shown in Figs.~\ref{t2n}
and \ref{c2n}. We then address the most significant question
concerning the limits on coupling and $\nu$ masses, when simultaneously
considered. 
\begin{figure}[h!]
\centering
\vskip .2truecm
\includegraphics[height=11.cm,angle=0]{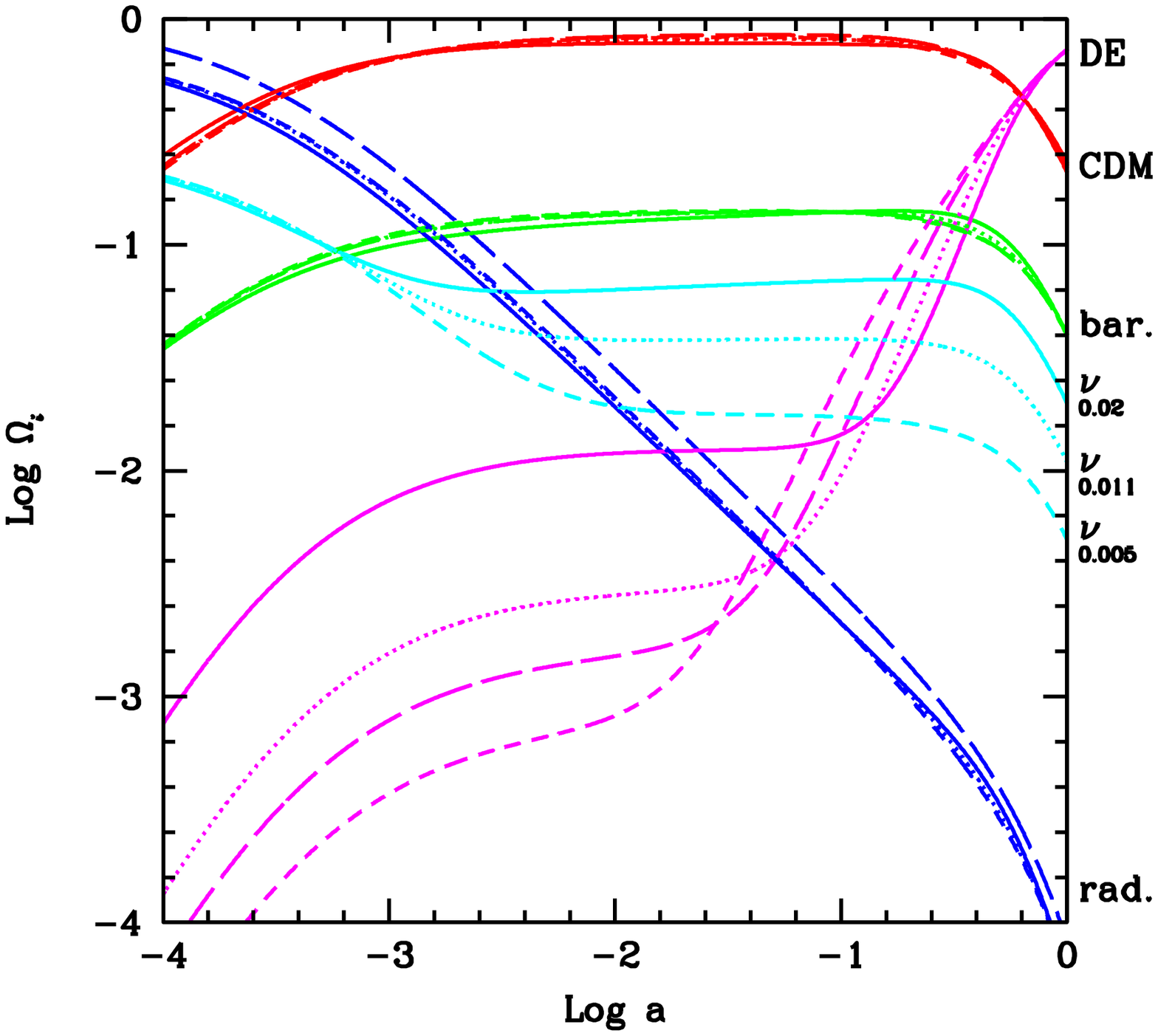}
\caption{\textsl{Density parameters for CDM, hot DM ($\nu$'s), DE
and radiation in models with $\Omega_\nu$ and $\beta$ taking the
values 0.005--0.049 (short dashed), 0--0.07 (long dashed), 0.011--0.1
(dotted), 0.02--0.21 (solid), respectively. In the last case, the DE
plateau, extending up to the equality redshift, occurring slightly
above $z \sim 10^3$, shows a DE density keeping $\sim 1/20$ of CDM.}}
\label{coinci}
\vskip +.1truecm
\end{figure}

This question should be carefully addressed by using MonteCarlo
techniques and considering all available observational constraints.
Unfortunately, to do so, we should widen the usual parameter space, by
adding 3 extra degrees of freedom: the coupling parameter $\beta$,
neutrino mass, and the energy scale $\Lambda$ in the SUGRA model (or
another equivalent parameter, in the same or in another dynamical DE
potential).

This is among the reasons that led previous authors to perform a
preliminary test by using a Fisher matrix technique. As is known, a FM
analysis assumes that a given cosmology has a top likelihood and
explores its neighbours. We can imagine, {\it e.g.} that the data
favor $\beta\neq 0$ in association with a significant $\nu$ mass.  A
FM analysis would hardly tell us that. This calls for a MCMC analysis
that we reserve for future work.

Taking FM outputs at face value leads to state that models with
$\Omega_\nu \lesssim 0.022$ and $\beta \lesssim 0.22$ are
observationally consistent with a $\Omega_\nu = 0$ and $\beta = 0$
model. 

Models with cDE were initially considered to overcome the {\it
coincidence} problem, in the presence of DE. To achieve this aim, one
has to accept that DE is characterized by two different scales. For a
coupled SUGRA cosmology like the one explored here, they are about the
EW and the Planck scales. No {\it ad--hoc} scale is then apparently
introduced, but the whole framework appears somehow artificial and
demands for a more basic scheme, to reduce its complexity. An example
is the {\it double--axion} model \cite{DA}, which however leads to
features different from the ones considered here.

Still working at a phenomenological level, in Figure \ref{coinci}, we
however show the scale dependence of the density parameters, for
various models with different $\sum m_\nu$ and $\beta$.

In the usual case, with negligible $m_\nu$, a CDM--DE coupling
compatible with data hardly eases the coincidence problem.  Such
easing is represented by the {\it plateau} in the $\Omega_{de}$ curve,
whose proportions are then almost insignificant. This does not mean
that $\beta \neq 0$ is not to be considered among the possible degrees
of freedom; {\it e.g.}, in \cite{vergani} it is shown that a cosmology
with $\beta$ as small as $ \sim 0.05$, if inspected assuming $\beta
\equiv 0$, can yield wrong values for some cosmic parameters,
including $\omega_{oc}$. Figure \ref{coinci} however indicates that,
when $\beta \sim 0.2$ is recovered, in the presence of suitably
massive $\nu$'s, a significant {\it plateau} is present and DE density
keeps at the level $\sim 1$--$2\, \%$ of the critical density up to $z
\sim 10^3$.

\ack 
Thanks are due to R, Mainini for discussions. LPLC was supported by
NASA grant NNX07AH59G and Planck subcontract 1290790 for this work,
and would like to thank the Physics Department G.~Occhialini, for
hospitality.

\appendix

\section{Fluctuation evolution in models with dynamical and coupled DE}

{\it Uncoupled DE equations} --
Let 
$$
 \Phi(\tau,k) = \phi(\tau) + \varphi(\tau,k)
\eqno (A1)
$$
be the dynamical DE field. In eq.~(A1) the background component
$\phi$, independent from space coordinates, and its fluctuations
$\varphi$ are outlined. The dependence on spatial coordinates is also
Fourier--transformed, so to have a $\varphi$ field dependent on the
time $\tau$ and the wavenumber $k$.

In the absence of CDM--DE coupling, the background equation
$$
\ddot \phi + 2(\dot a/a) \dot \phi + a^2 V'(\phi) = 0
\eqno (A2)
$$
is known to hold, together with the equation
$$
\ddot \varphi + 2(\dot a/a) \dot \varphi + \dot \phi \dot h /2
+ k^2 \phi + a^2 V''(\phi) \varphi = 0
\eqno (A3)
$$
for its fluctuations. Here $\dot h$ is the usual variable describing
the gravitational field due to density fluctuations in the synchronous
gauge.

Aside of them, the equations
$$
\dot \rho_c + 3(\dot a/a) \rho_c = 0
\eqno (A4)
$$
$$
\dot \delta_c + k v_c + \dot h/2 = 0
\eqno (A5a)
$$
$$
\dot v_c + (\dot a/a) v_c = 0~.
\eqno (A5b)
$$
will hold, in a generic synchronous gauge, for the CDM density
$\rho_c$ and its fluctuations $\delta_c$. Here $v_c$ is the (gauge
dependent) {\it velocity field} in CDM.

The presence of dynamical DE clearly implies that the equation
fulfilled by the scale factor is also modified into
$$
\dot a^2 = a^4 H_o^2 \left[ \Omega_{o\gamma}/a^4 +
(\Omega_{ob}+\Omega_{oc})/a^3 + (\rho_k+V)/\rho_{o,cr} +
\rho_\nu/\rho_{o,cr}\right]~.
\eqno (A6)
$$
The symbols $H_o$, $\Omega_{o\gamma}$, $\Omega_{oc}$, $\Omega_{ob}$
have their obvious meaning; $\rho_{o,cr}$ is the present critical
energy density, while
$$
\rho_k = \dot \phi^2/2a^2
\eqno (A7)
$$ 
is the kinetic energy density of the DE field.
Finally
$$
\rho_\nu = {{\cal N}_\nu \over 2\pi^2} T_\nu^4 \int_0^\infty dx ~x^3 
{\epsilon(m_\nu/xT_\nu) \over \exp \epsilon(m_\nu/xT_\nu) + 1}
~~~~
{\rm with}
~~~
\epsilon(\mu) = \sqrt{1+\mu^2}
$$ 
is the energy density of $\nu$'s with mass $m_\nu$ and $\cal N$$_\nu$
spin states.

This equation shall not be modified by the presence of coupling, as
well as the equations for gravitational field fluctuations, to whose
source $\varphi$ contributes. We shall omit these last equations.

\vskip .3truecm

{\it Coupled DE equations} -- In the presence of a constant CDM--DE
coupling, eqs.~(A2)--(A5) are modified so that their r.h.s.'s no
longer vanish. More in detail~:
$$
(A2)-(A4)~~ \to ~~~~~~~~~
\ddot \phi + .... = Ca^2 \rho_c~,~~~~~
\dot \rho_c + .... = -C \rho_c \dot \phi
$$
$$
(A3) ~~~~~~~~~~~~ \to ~~~~~~~~~
\ddot \varphi + ..... = C a^2 \rho_c \delta_c
~~~~~~~~~~~~~~~~~~~~~~~~~~~
$$
$$
(A5) ~~~~~~~~~~~~ \to ~~~~~~~~~~
\dot \delta_c + ..... = -C \dot \varphi~,~~~
\dot v_c + ...... = -kC \varphi
$$

All other dynamical equations keep unmodified.


\begin{thebibliography}{50}

\bibitem{solar} 
Q. R.Ahmad {\it et al.}, \PRL {\bf 89}, 011301 (2002); 
S. N.Ahmed {\it et al.}, \PRL {\bf 92}, 181301 (2004).

\bibitem{reactor} 
K. Eguchi {\it et al.}, \PRL {\bf 90}, 021802h (2003); 
T. Araki {\it et al.}, \PRL {\bf 94}, 081801 (2005).

\bibitem{atmo} 
W. W. Allison {\it et al.}, \PL B{\bf 449}, 137 (1999); 
M. Ambrosio {\it et al.}, \PL B{\bf 517}, 59 (2001).

\bibitem{beam}
M. H. Ahn {\it et al.}, \PRL {\bf 90}, 041801h (2003);
D. G. Michael {\it et al.}, \PRL {\bf 97}, 191801 (2006).

\bibitem{bono1} 
R. Valdarnini \& S. Bonometto, A\&A {\bf 146}, 2 235 (1985); 
R. Valdarnini \& S. Bonometto, \APJ {\bf 299}, L71 (1985).

\bibitem{bib1} 
A. G. Riess, R. P. Kirshner, B. P. Schmidt, S. Jha, {\it et al.}, \APJ {\bf 116}, 1009 (1998); 
S. Perlmutter, G. Aldering, G. Goldhaber, {\it et al.}, \APJ {\bf 517}, 565 (1999); 
A. G. Riess, L.G. Strolger, J. Tonry, {\it et al.}, \APJ {\bf 607}, 665 (2004).

\bibitem{bib2} 
P. de Bernardis, P. Ade, J. Bock, {\it et al.}, Nat. {\bf 404}, 955 (2000); 
S. Padin, J. Cartwright, B. Mason, {\it et al.}, \APJ {\bf 549}, L1 (2001); 
J. Kovac, E. Leitch, c. Pryke, {\it et al.}, Nat. {\bf 420}, 772 (2002); 
P. Scott, P.  Carreira, K. Cleary, {\it et al.}, \MNRAS {\bf 341}, 1076 (2003);
D. Spergel, R. Bean, Dor\`e {\it et al.}, \APJ Suppl. {\bf 170}, 377 (2007).

\bibitem{bib3} 
M. Colless, G. Dalton, S. Maddox, {\it et al.}, \MNRAS {\bf 329}, 1039 (2001);
M. Colless, B. Peterson, C. Jackson, {\it et al.}, Preprint astro--ph/0306581; 
J. Loveday (the SDSS collaboration), \CP {\bf 43}, 437 (2002); 
M. Tegmark, M. Blanton, M. Strauss, {\it et al.}, \APJ {\bf 606}, 702 (2004); 
J. Adelman--McCarthy, M. Agueros, S. Allam, {\it et al.}, \APJ Suppl. {\bf 162}, 38 (2004).

\bibitem{nu1} 
A. Dolgov, \PR {\bf 370}, 333 (2002); 
O. Elgaroy \& O. Lahav, \NJP {\bf 7}, 61t (2005); 
J. Lesgourgues \& S. Pastor, \PR {\bf 429}, 307 (2006).

\bibitem{fogli}
G.~L.~Fogli {\it et al.}, Phys.\ Rev.\  D {\bf 78}, 033010 (2008);
G.~L.~Fogli {\it et al.}, Phys.\ Rev.\  D {\bf 75}, 053001 (2007).

\bibitem{bib30} 
E. Komatsu {\it et al.} [WMAP Collaboration], Preprint arXiv:0803.0547 [astro-ph].

\bibitem{others1}
A.~Goobar, S.~Hannestad, E.~Mortsell \& H.~Tu, JCAP {\bf 0606}, 019 (2006);
A.~G.~Sanchez \& S.~Cole, \MNRAS, {\bf 385}, 830 (2008).

\bibitem{bib33} 
W. Percival {\it et al.}, Monthly Notices of the Royal Astronomical Society {\bf 327}, 1297 (2001).

\bibitem{bib34} 
D. G. York {\it et al.}, Astronomical Journal {\bf 120}, 1579 (2000); 
C. Stoughton {\it et al.}, Astronomical Journal {\bf 126}, 485 (2002);
K. Abazajian {\it et al.}, Astronomical Journal {\bf 123}, 2081 (2003).

\bibitem{others2}
U.~Seljak {\it et al.}  [SDSS Collaboration], \PR D {\bf 71}, 103515 (2005);
U.~Seljak, A.~Slosar and P.~McDonald, JCAP {\bf 0610}, 014 (2006).

\bibitem{nulens}
K.~N.~Abazajian \& S.~Dodelson, \PRL  {\bf 91}, 041301 (2003);
S.~Hannestad, H.~Tu \& Y.~Y.~Y.~Wong, JCAP {\bf 0606}, 025 (2006);
T.~D.~Kitching, A.~F.~Heavens, L.~Verde, P.~Serra \& A.~Melchiorri, \PR D {\bf 77}, 103008 (2008).

\bibitem{coupling}
Wetterich C., A\&A {\bf 301}, 321 (1995); 
Amendola L., \PR D{\bf 62}, 043511 (2000).

\bibitem{claudia}
L.~Amendola and C.~Quercellini, \PR D {\bf 68}, 023514 (2003).

\bibitem{maccio}
Maccio' A. V., Quercellini C., Mainini R., Amendola L., Bonometto S. A., \PR D{\bf 69}, 123516 (2004).

\bibitem{interaction} 
J.~Ellis, S.~Kalara, K.~A.~Olive \& C.~Wetterich, \PL B228, 264 (1989); 
M. Gasperini, F. Piazza \& G. Veneziano, \PR D65, 023508 (2002); 
D.~Comelli, M.~Pietroni and A.~Riotto, \PL B {\bf 571} 115 (2003); 
L.P. Chimento, A.S. Jakubi, D. Pavon \& W. Zimdahl, \PR D67, 083513 (2003);
Z. K. Guo, N. Ohta and S. Tsujikawa, \PR D {\bf 76} 023508 (2007); 
E. Abdalla, L. R. W. Abramo, L. J. Sodre and B. Wang, {\it Preprint} arXiv:0710.1198 [astro-ph]; 
M. Manera and D. F. Mota, \MNRAS {\bf 371} 1373 (2006).

\bibitem{chamaleon}
J.~Khoury and A.~Weltman , \PRL {\bf 93} (2004) 17110;
S.S.~Gubser and J.Khoury, \PR D {\bf 70} (2004) 104001;
P.~Brax et al., \PR D {\bf 70} (2004) 123518;
D.~F.~Mota and D.~J.~Shaw, \PRL {\bf 97}, 151102 (2006);
P.~Brax, C.~van de Bruck, A.~C.~Davis, D.~F.~Mota and D.~J.~Shaw, \PR D {\bf 76}, 124034 (2007).

\bibitem{Hannestad:2005gj}
S.~Hannestad, \PRL {\bf 95}, 221301 (2005).

\bibitem{DeLaMacorra:2006tu}
 A.~De La Macorra, A.~Melchiorri, P.~Serra \& R.~Bean, Astropart.\ Phys.\  {\bf 27}, 406 (2007).

\bibitem{bib6} 
L. Colombo \& M. Gervasi, JCAP {\bf 10}, 001 (2006).

\bibitem{bib5} 
P. Brax \& J. Martin, \PL B{\bf 468}, 40 (1999); 
P. Brax \& J. Martin, \PR D{\bf 61}, 103502 (2000);
P. Brax, J. Martin \& A. Riazuelo, \PR D{\bf 62}, 103505 (2000).
 
\bibitem{darmour1}
Damour T., Gibbons G. W. \&  Gundlach C., 1990,  Phys.Rev., L64, 123D
Damour T. \&  Gundlach C., 1991,  Phys.Rev., D43, 3873

\bibitem{brans}  
L.~Amendola, \PR  D {\bf 60}, 043501 (1999);
V.~Pettorino and C.~Baccigalupi, \PR  D {\bf 77}, 103003 (2008)

\bibitem{camb}
http://www.camb.info/

\bibitem{atv} L.~Amendola \& D.~Tocchi--Valentini, \PR D {\bf 66},
043528 (2002)

\bibitem{Fisher:1935}
R.~A.~Fisher, J. Roy. Statist. Soc. {\bf 98}, 39  (1935).

\bibitem{Sivia:1996} 
D.~S.~Sivia, {\it Data analysis: A Bayesian Tutorial}, Oxford Univ. Press (Oxford, 1996).

\bibitem{Tegmark:1996bz} 
M.~Tegmark, A.~Taylor \& A.~Heavens, \APJ, {\bf 480}, 22 (1997).

\bibitem{bib32} 
Planck Blue Book: http://www.rssd.esa.int/SA/PLANCK/docs/Bluebook-ESA-SCI(2005)1 V2.pdf

\bibitem{Neyrinck:2006zi} 
M.~C.~Neyrinck \& I.~Szapudi, \MNRAS, {\bf 375}, L51 (2007).

\bibitem{Scoccimarro1999}
R.~Scoccimarro, M.~Zaldarriaga \& L.~Hui, \APJ,{\bf 527}, 1 (1999).

\bibitem{Hamilton2006} 
A.~J.~S.~Hamilton, C.~D.~Rimes \& R.~Scoccimarro, \MNRAS,{\bf 371}, 1188 (2006)

\bibitem{couplconstr}
G.~Olivares, F.~Atrio-Barandela and D.~Pavon, \PR D {\bf 77}, 063513 (2008).

\bibitem{DA} 
R.~Mainini, S.A.~Bonometto, \PRL  {\bf 93}, 121301 (2004);
R.~Mainini, L.P.L.~Colombo, S.A.~Bonometto, Astrophys.\ J.\  {\bf 632}, 691 (2005);
R.~Mainini, S.A.~Bonometto, JCAP {\bf 0709}, 017 (2007).

\bibitem{vergani}
G.~La Vacca, L.~P.~L.~Colombo, L.~Vergani \& S.~A.~Bonometto, arXiv:0804.0285 [astro-ph].

\end{thebibliography}
\end{document}